\documentclass[epj]{svjour}
\usepackage[dvips]{graphics}
\begin{document}
\title{An investigation of the quantum $J_1-J_2-J_3$ model on the
honeycomb lattice}
\author{
J.B. Fouet, P. Sindzingre, C. Lhuillier }
\institute{Laboratoire de Physique Th{\'e}orique des
  Liquides, Universit{\'e} P. et M. Curie, case 121, 4 Place Jussieu, 75252
  Paris Cedex. UMR 7600 of CNRS.}
\date{\today}

\bibliographystyle{prsty}

\abstract{
We have investigated  the  quantum 
$J_1-J_2-J_3$ model on the honeycomb lattice
with exact diagonalizations and linear spin-wave calculations
for selected values of $J_{2}/J_{1}$, $J_{3}/J_{1}$ 
and antiferromagnetic ($J_{1}>0$) or ferromagnetic ($J_{1}<0$)
nearest neighbor interactions.
We found a variety of quantum effects:
"order by disorder" selection of a N{\'e}el ordered ground-state,
good candidates for 
non-classical ground-states with dimer long range order 
or spin-liquid  like.
The purely antiferromagnetic Heisenberg model is confirmed to 
be N{\'e}el ordered.
Comparing these results with those observed on the square
and triangular lattices, we enumerate some conjectures on the nature of the
quantum phases in the isotropic models.
}

\maketitle
\begin{section}{Introduction} 
\label{sec:intro}
Frustrated quantum antiferromagnetic (AF) spin systems 
on low dimension (D) lattices 
have attracted a great deal of interest in recent years. 
Quantum fluctuations, largest for small values of the spin $S$
of the magnetic ions, low D and small coordination of the lattice,
are expected  to lead to novel magnetic behaviors.
Their effects, have been preeminently seen in 1D. 
They have been investigated on a few 2D systems.
The most studied models are the AF Heisenberg model 
on the triangular~\cite{blp92,capriotti99}
or kagom{\'e} lattice~\cite{smlbpwe00}
which are geometrically frustrated systems, 
the $J_1-J_2$ model on the square lattice~\cite{capriotti00,singh99,kos99}
where frustration is introduced by  2nd neighbor interaction,
the $J_1-J_2$ model~\cite{cj92,k93,lblp95a} 
and the multi-spin exchange model  (MSE)~\cite{mblw98},
on the triangular lattice.

Less studied~\cite{oitmaa78,reger89,oitmaa92,weihong91},
spin models on the honeycomb lattice deserve attention 
due to the special properties of the lattice
and since there are experimental realizations.
A first feature of the lattice is that,
like the square lattice, it is not geometrically frustrated 
for AF nearest neighbor interactions but has lower coordination. 
Thus quantum fluctuations are expected to be larger than for
the square lattice.
For this reason the spin-1/2 Heisenberg antiferromagnet 
on the honeycomb lattice,
has been studied theoretically by various methods
~\cite{oitmaa78,reger89,oitmaa92,weihong91}
which all predicted that N{\'e}el long range order (LRO) subsists but with  an 
order parameter smaller than for the square  lattice case.
This also motivated a Schwinger-boson study of 
the effects of frustrating second neighbor interactions in the
$J_1-J_2$ model~\cite{mattsson94}.

A major incentive to study frustrated magnets on the honeycomb lattice
is the availability of experimental data in the family of compounds 
${\rm BaM_2(XO_4)_2 }$ (M= Co, Ni; X= P, As)
obtained, some years ago, by Regnault and Rossat-Mignod~\cite{regnault-1}.
The magnetic ions M have small spins (it is supposed to be $S=1/2$ for the Co
oxide and  $S=1$ for Ni),
disposed in weakly coupled layers where they sit on a honeycomb lattice.
The simplest model relevant to these quasi 2D compounds
is a $J_1-J_2-J_3$ model on a honeycomb lattice 
with first, second and third neighbor interactions
and either on site if ($S=1$) or XXZ (if $S=1/2$) anisotropy.

So far the $J_1-J_2-J_3$ model was only investigated within first order
linear spin-wave  
theory (LSW)~\cite{regnault-1,rastel79}, and to our knowledge the renormalization
of the order parameter by quantum fluctuations has not be calculated even in this
simplest approach. The experimental results motivated us to do this
calculation in the large S limit and then attack the  $S=1/2$ problem with
exact diagonalizations (ED).

In this paper, as a first step, we consider the case of purely
isotropic interactions.
The Hamiltonian of the model reads:
\begin{equation}
{\cal H} =  J_1 \sum_{<i,j>_{1}} {\bf S}_i.{\bf S}_j
+ J_2 \sum_{<i,k>_{2}} {\bf S}_i.{\bf S}_k
+ J_3 \sum_{<i,k>_{3}} {\bf S}_i.{\bf S}_k
\label{eq-Heij3}
\end{equation}
where  the first, second and third sums run on the first, second
and third neighbor pairs of spins, respectively (see Fig.\ref{fig-hon-latt}). 
The coupling constants $J_i$ can be either AF ($J_i>0$) 
or ferromagnetic ($J_i<0$).
Depending of the values of the parameters $J_i$,
this model displays various classical ground-states:
a collinear AF ground-state, two degenerate manifolds of
 planar  helimagnetic ground-states
with four or eight sublattices 
and a ferromagnetic ground-state.
The classical phase diagrams of the isotropic and anisotropic XXZ models
display only minor differences.
In particular  the classical  ground-states are the same
for the parameters believed to be relevant to ${\rm BaCo_2(AsO_4)_2 }$.
In this paper we concentrate on the quantum effects in the isotropic model.
The study of the quantum XXZ $J_1-J_2-J_3$ model
with parameters appropriate to ${\rm BaCo_2(AsO_4)_2 }$
will be presented in a separate paper~\cite{fouet00b}.

\begin{figure}
	\begin{center}
	\resizebox{8cm}{!}{
	\includegraphics{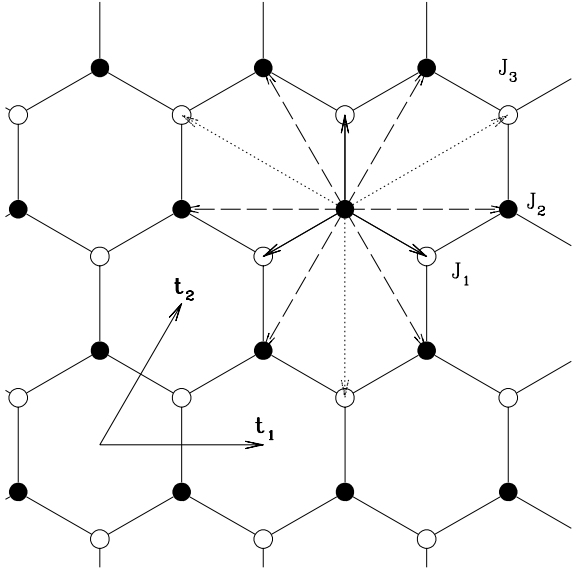}}
	\end{center}

    \caption[99]{ The honeycomb lattice. 
The full and empty circles differentiate the two sublattices.
${\bf t}_1$ and ${\bf t}_2$ are the two vectors of the triangular Bravais
lattice.
Top right: arrows show the interactions between a site and its
1th, 2th and 3th neighbors.
}
    \label{fig-hon-latt}
\end{figure}

We investigated quantum effects for selected values of the $J_i$
chosen to give a broad picture of the quantum effects encountered
in the model.
The restriction to isotropic interactions
limits the number of parameters and enable to separate 
the effects of anisotropy.
In addition to the $J_1-J_2$ models on the square and triangular lattices,
the present model may be compared with 
the $J_1-J_2-J_3$ model on the square lattice 
which has a similar variety of classical ground-states
and to which a few studies have been 
devoted~\cite{fkksrr90,moreo90,read-sachdev91,ceccatto93,ferrer93,ll96}.

This paper is divided into five parts.
In section II, we recall the classical phase diagram of the model,
obtained by Rastelli {\it et~al.}~\cite{rastel79}, 
and identify the degeneracies
of the classical ground-states not considered 
by these authors, we also discuss the stability of the first order spin-wave
approximation for these different phases.
The ED results for the case of antiferromagnetic and ferromagnetic
nearest neighbor coupling are presented in section III and IV respectively.
In section V we draw conclusions, and enumerate some conjectures
relative to the appearance of the various generic two-dimensional spin-liquids. 
We described in an appendix the various technical features specific to our
present ED calculations on different samples.

\end{section}

\begin{section}{Classical phase diagram and semi-classical deviations} 
\label{sec:2}

\subsection{ Planar ground-state configurations}
\label{sec:2.1}
The classical model was studied by Rastelli {\it et~al.}~\cite{rastel79}. 
They searched for planar or uniformly canted configurations 
minimizing the classical energy $E_{cl}$.
The former were found energetically favored over the latter.
They represent spiral configurations,
characterized by a wave-vector ${\bf Q}$.
The classical spin (of length $S$) sitting at
cell ${\bf R}$ of the triangular Bravais lattice
on sublattice $\alpha$ is given by:
\begin{equation}
{\bf S}_{{\bf R},\alpha}=
  S \large( \cos\left({\bf Q}.{\bf R}+\phi_{\alpha}\right) {\bf u} 
+   \sin\left({\bf Q}.{\bf R}+\phi_{\alpha}\right) {\bf v} 
   \large)
\label{eq-planar_sol}
\end{equation}
where 
${\bf u}$ and {\bf v} are two orthogonal unit vectors 
defining the plane of the spins,
$\phi_{\alpha}$ can be chosen to be zero on one sublattice and 
will be noted $\phi$ on the other. 
The set of spiral wave-vectors ${\bf Q}$ 
minimizing the classical energy will be noted $\{\bf Q\}$. 

The phase diagram  of planar solutions of this type  
is reproduced in Fig.\ref{fig-class-phas-a} 
($J_{1}>0$) and Fig.\ref{fig-class-phas-b} ($J_{1}<0$).
There is a mapping  between the two phase diagrams:
the transformation $J_{1}\to -J_{1}$,
$J_{3}\to -J_{3}$ and ${\bf S}_i \to -{\bf S}_{i}$
for $i \in $  on the black triangular sublattice of Fig.1
leaves the Hamiltonian unchanged, and maps the ground-state for
$J_{1}>0$ on that for $J_{1}<0$.

\begin{figure}
	\begin{center}
	\resizebox{8cm}{!}{
	\includegraphics{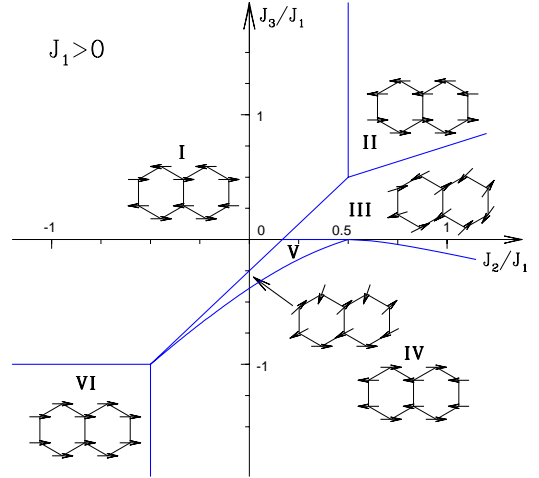}}
	\end{center}

    \caption[99]{ Classical phase diagram for 
antiferromagnetic nearest neighbor interactions. In the $T=0$ classical approximation
 regions II and IV have in fact a degenerate manifold of non-planar
ground-states. Thermal fluctuations  or quantum fluctuations  do select the
collinear configurations shown in this figure. }
    \label{fig-class-phas-a}
\end{figure}

There are six regions in each phase diagram:
four collinear phases (I,II,IV,VI) and two spiral ones (III,V).
In the collinear regions I and VI, the wave-vector of the magnetic order 
is ${\bf Q}=0$, 
whereas in regions II and IV, $\{\bf Q\} = \{\bf K_i\}$, 
where ${ {\bf K}_{i} }$ are the three inequivalent  middles of edges
of the Brillouin zone (see Fig.\ref{fig-BZ}).
The phases are: In I, $\phi=\pi$ ($0$) if $J_1>0$ ($J_1<0$),
in VI, $\phi=0$ ($\pi$) if $J_1>0$ ($J_1<0$),
in II $\phi=\pi$ ($0$) if $J_1>0$ ($J_1<0$),
in IV $\phi=0$ ($\pi$) if $J_1>0$ ($J_1<0$). 

The separation lines I-III, I-V, II-III, IV-V represent continuous
transitions; the others are first order phase transitions.

\begin{figure}
	\begin{center}
	\resizebox{8cm}{!}{
	\includegraphics{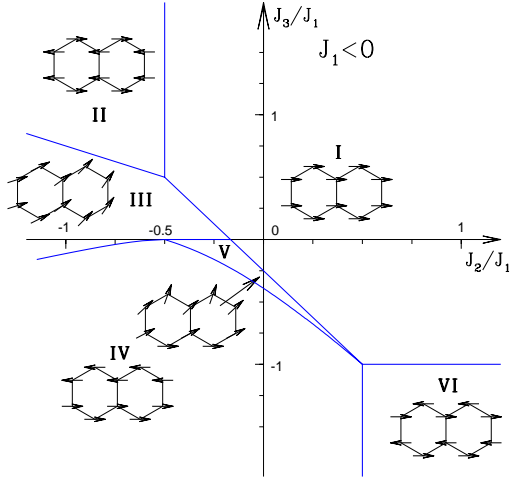}}
	\end{center}

    \caption[99]{ Classical phase diagram for
ferromagnetic nearest neighbor interactions. Same comments on regions II and IV
as in Fig.\ref{fig-class-phas-a}}
    \label{fig-class-phas-b}
\end{figure}

\subsection{Non planar ground-states manifolds}
\label{sec:2.2}
Ansatz (Eq.\ref{eq-planar_sol}) is usually generic to
find all allowed classical ground-states~\cite{villain77,bertaut63}.
It assumes that, up to the trivial degeneracy  associated to a  global spin
rotation,
the ground-state is unique or exhibits at most a discrete 
degeneracy.
It is valid if a linear combination of the different ${\bf Q}$ modes
of the same set $\{\bf Q\}$,
 can be excluded  as violating the constraint 
$|{\bf S}_{{\bf R},\alpha}|=S$ 
on every site.
Exceptions occur for special sets $\{\bf Q\}$, in particular if 
${\bf Q}$ is half or one fourth of a reciprocal lattice vector 
${\bf G}$~\cite{villain77}.
This is the case in region II and IV 
where  $\{\bf Q\} = \{\bf G/2\}=\{\bf K_i\}$ (see Fig.6).
Here, the linear combination of three ${\bf K}_{i}$ solutions
\begin{equation}
{\bf S}_{{\bf R},\alpha}= \sum_{i=1}^{3}
 S \cos\left({\bf K}_{i}.{\bf R}+\phi_{\alpha}\right)
 {\bf u}_{i}
\label{eq-3q-sol}
\end{equation}
with unnormalized ${\bf u}_{i}$, is submitted to 
the constraints  ${\bf u}_{i}.{\bf u}_{j}=\delta_{i,j}$
and $\sum {\bf u}_{i}^{2}=1$.
The ground-state is a two dimensional manifold 
continuously connecting the three ${\bf K}_{i}$ solutions
(there are nine degrees of freedom for choosing the three ${\bf u}_{i}$,
minus three for global rotations of the spins and four constraints).
This gives birth to the non planar ground-states manifolds described below.

 In regions IV for $J_1>0$ or  II for $J_1<0$ (
$\phi=0$),
the ground-state manifold is the set of 
four-sublattice ordered solutions 
such as ${\bf S}_{A}+{\bf S}_{B}+{\bf S}_{C}+{\bf S}_{D}=0$.
This could be seen directly from the expression of the
classical energy of these  configurations:
\begin{eqnarray}
E_{cl} &=&
\frac{2}{N} \left(J_1+ 2J_2 \right)
\left({\bf S}_{A}+{\bf S}_{B}+{\bf S}_{C}+{\bf S}_{D} \right)^{2}
\\ \nonumber
&-& \frac{2}{N} \left(J_1+ 2J_2 -3J_3 \right)
\left({\bf S}_{A}^{2}+{\bf S}_{B}^{2}+{\bf S}_{C}^{2}+{\bf S}_{D}^{2}\right).
\label{eq-4_sr_af}
\end{eqnarray}

In this equation, N represents the total number of spins of the sample.
The generic four-sublattice configurations are shown in
Fig.\ref{fig-4_sr_af}, as well as the three collinear configurations,
which appear as special cases of it, with  ${\bf S}_{A}={\bf S}_{B}= -{\bf
S}_{C}=-{\bf S}_{D}$ and the two other combinations of parallel spins.
\begin{figure}
	\begin{center}
	\resizebox{8cm}{!}{
	\includegraphics{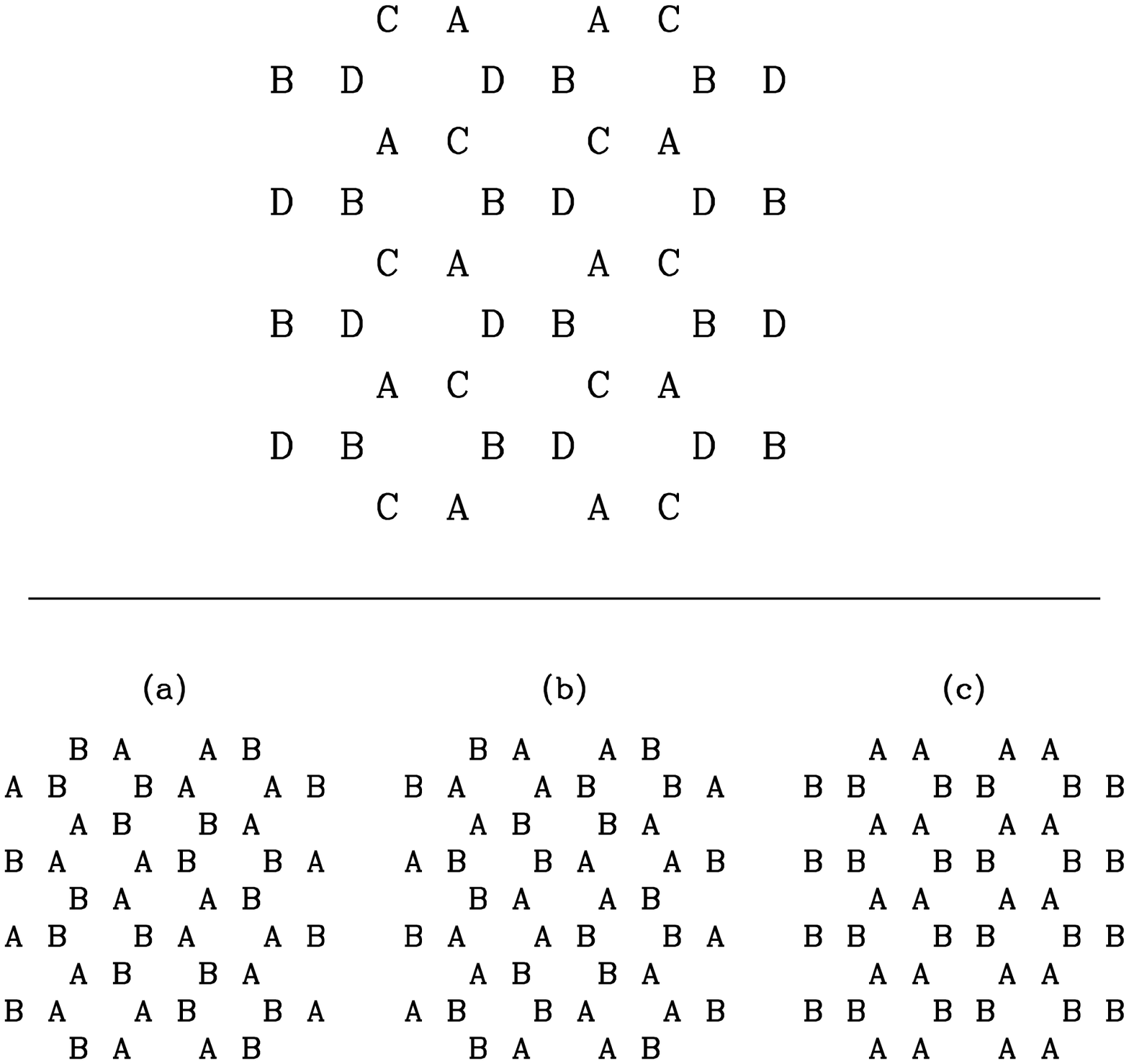}}
	\end{center}

    \caption[99]{ 
Top: four-sublattice classical ground-state in region IV
on Fig.\ref{fig-class-phas-a} for 
 antiferromagnetic first neighbor coupling ($J_1>0$).
Bottom:  the collinear solutions with the three possible arrangements
(in this case, classical spins in sublattices A and B are antiparallel).} 
    \label{fig-4_sr_af}
\end{figure}
The situation is reminiscent of the $J_1-J_2$ model on the 
triangular lattice~\cite{cj92,k93,lblp95a} with $1/8< J_2/J_1 <1$.

In regions IV for $J_1<0$ or in II for $J_1>0$ ($\phi=\pi$), 
the ground-state manifold is the  set of 
eight-sublattice solutions shown in Fig.\ref{fig-4_sr_fe}  
where sublattices labelled by the same letter 
are paired (partner sublattice is over-headed by a bar) and
\begin{equation}
{\bf S}_{\bar{\alpha}}=-{\bf S}_{\alpha} 
\end{equation}
for $\alpha \in \{A,B,C,D\}$
and
\begin{equation}
{\bf S}_{A}+{\bf S}_{B}+{\bf S}_{C}+{\bf S}_{D}=0.
\end{equation}
This minimizes the classical energy:

\begin{eqnarray}
\nonumber
E_{cl}  &=&
\frac{8}{N} J_2
\large({\bf S}_{A}+{\bf S}_{B}+{\bf S}_{C}+{\bf S}_{D} 
\\ \nopagebreak \nonumber & & 
\;\;\;\;\;\;  
+ {\bf S}_{\bar{A}}+{\bf S}_{\bar{B}}+{\bf S}_{\bar{C}}+{\bf S}_{\bar{D}}
\large)^{2}
\\ \nopagebreak \nonumber &+& \frac{8}{N} \left(J_1 - 2J_2 \right)
\large( {\bf S}_{A}+{\bf S}_{B}+{\bf S}_{C}+{\bf S}_{D} \large)
\\ \nopagebreak \nonumber & & 
\;\;\;\;  \;\;\;\;  \;\;\;\;  \;\;\;\;  
\left(
    {\bf S}_{\bar{A}}+{\bf S}_{\bar{B}}+{\bf S}_{\bar{C}}+{\bf S}_{\bar{D}}
\right)
\\ \nonumber  
&+& \frac{4}{N} \left( 3J_3 - J_1 \right)
 \large[
  \left({\bf S}_{A}+{\bf S}_{\bar{A}} \right)^{2}
+ \left({\bf S}_{B}+{\bf S}_{\bar{B}} \right)^{2}
\\ \nonumber & & 
\;\;\;\;  \;\;\;\; \;\;\;\;  \;\;\;\; \;\;\;\;  
+ \left({\bf S}_{C}+{\bf S}_{\bar{C}} \right)^{2}
+ \left({\bf S}_{D}+{\bf S}_{\bar{D}} \right)^{2}
 \large]
\\ \nonumber 
&-& \frac{4}{N} \left(-J_1+ 2J_2 +3J_3 \right)
 \large(
{\bf S}_{A}^{2}+{\bf S}_{B}^{2}+{\bf S}_{C}^{2}+{\bf S}_{D}^{2}
\\  & &
\;\;\;\;  \;\;\;\;  \;\;\;\;  \;\;\;\;  \;\;\;\;  \;\;\;\;  
    + {\bf S}_{\bar{A}}^{2}+{\bf S}_{\bar{B}}^{2}
     +{\bf S}_{\bar{C}}^{2}+{\bf S}_{\bar{D}}^{2}
 \large).
\label{eq-4_sr_fe}
\end{eqnarray}

\begin{figure}
	\begin{center}
	\resizebox{8cm}{!}{
	\includegraphics{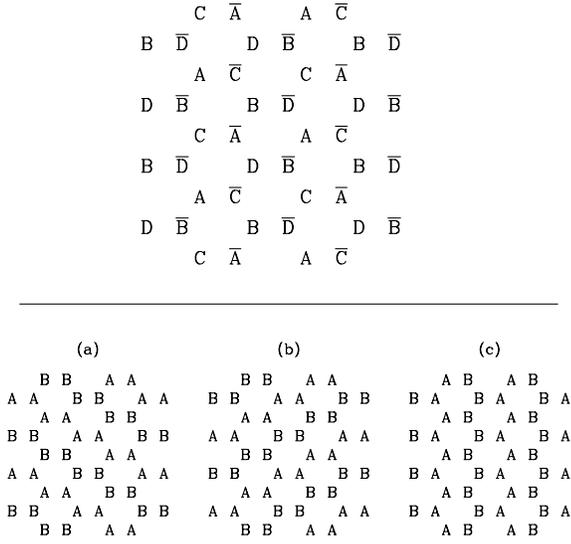}}
	\end{center}
    \caption[99]{ Same as  Fig.\ref{fig-4_sr_af}) but for ferromagnetic
first neighbor coupling ($J_1<0$), region IV of Fig.\ref{fig-class-phas-b}.
}
    \label{fig-4_sr_fe}
\end{figure}                                                            

It is highly probable that thermal fluctuations will stabilize the collinear
solutions as it does in similar situations on the square and triangular
lattice. We will show below that quantum fluctuations indeed do it.

Continuous degeneracy of the ground-state also occurs 
when ${\bf Q}={\bf G}/4$ in III  and in V  but
since this happens only on lines (for instance if $J_2=0.5$ in III) 
and not in full regions, 
we shall  skip their description which is not essential to our present
goal.
The transition line III-V between the two spiral regions is very special.
It has an infinite degeneracy of spiral ground-states corresponding to:
\begin{equation}
\cos({\bf Q}.{\bf t}_{1}) + \cos({\bf Q}.{\bf t}_{2}) + \cos({\bf Q}.({\bf t}_{1}
- {\bf t}_{2})) = 1/8J_2^2-3/2
\label{eq-III-V_q}
\end{equation}
The lines of ${\bf Q}$ solutions of Eq.\ref{eq-III-V_q} are 
shown in Fig.\ref{fig-BZ} for $J_2=0.2,0.4,0.5$.
\begin{figure}
	\begin{center}
	\resizebox{8cm}{!}{
	\includegraphics{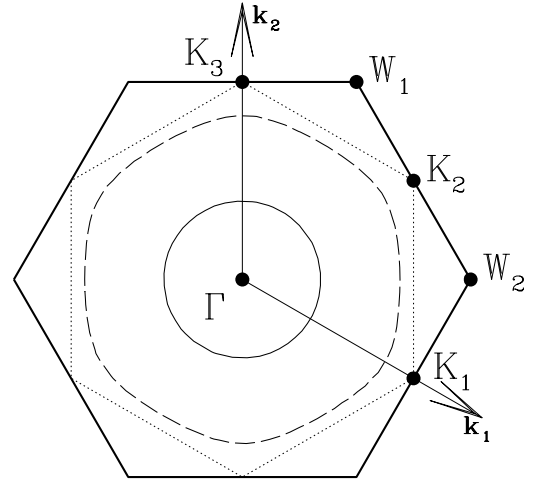}}
	\end{center}
    \caption[99]{ Brillouin zone of the triangular Bravais lattice.
 Light solid, dashed and dotted lines  are the solutions of Eq.8, for $J_2=0.2, 0.4,0.5$ respectively.
}
    \label{fig-BZ}
\end{figure}

\subsection{Stability of the quasi-classical phase diagram in the large S
limit:  LSW results}
\label{sec:2.3}

The renormalization of the order parameter $m^{\dag}$ in the
first order spin-wave approximation is already large in the pure model
($J_1=1,J_2=0, J_3=0$): $m^{\dag} \approx 0.48$ 
i.e. a value reduced to $\approx 48\%$ of its classical value
 (Fig.\ref{fig-op-af}).

\begin{figure}
	\begin{center}
	\resizebox{8cm}{!}{
	\includegraphics{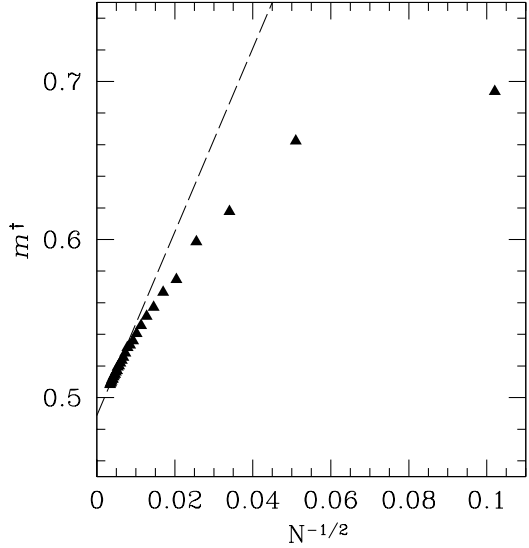}}
	\end{center}
    \caption[99]{LSW values (triangles) of the order parameter $m^{\dag}$
for the purely AF Heisenberg model ($J_1=1$, $J_2=0$,$J_3=0$),
as a function of $N^{1/2}$.
The asymptotic behavior $\sim N^{1/2}$ (dashed line fitted to the LSW values)
is only reached for quite large samples, much
larger than the sizes studied in exact diagonalizations.
}
    \label{fig-op-af}
\end{figure}

For $J_1>0$, the interplay
of quantum fluctuations and frustration quickly destroys N{\'e}el LRO:
$m^{\dag}$ goes to zero for $J_2\approx 0.1$ or $J_3\approx
-0.1$. The helimagnetic LRO of zone V disappears: 
near the $J_3=0$ axis, this is mainly due to the large classical
degeneracy of the ground-state  and near the $J_2=0$ axis, 
the main cause is the
vanishing of the spin-wave velocity at the point $J_1=1, J_2=0,
J_3=0.25$. The zone V being very small, we conclude  
that the N\'eel helimagnetic ground-state does not survive in region V for antiferromagnetic $J_1$.

For $J_1<0$ the ground-state is ferromagnetic in zone I of
figure \ref{fig-class-phas-b}. The ferromagnetic state is an exact
eigenstate of the hamiltonian, quantum fluctuations don't
destroy it. However, the classical degeneracy on the boundary between zones III and V
implies a whole branch of soft modes and a disappearance of
the helimagnetic LRO, on and near this line (as in the AF $J_1$ case).
The main difference with the AF first neighbor case is the persistence
of N{\'e}el order in zone V near the $J_1=-1, J_2=0, J_3=0.25$
point and in the vicinity of the $J_2=0$ axis.

The nature of the quantum spin-$1/2$ phase for small  $J_2$ and  $J_3$
appears an open problem that we will now attack with the help of
exact diagonalizations (ED).

Exact diagonalizations were performed on 
samples of $N=18,24,26,28,30,32$ sites with appropriate boundary conditions 
(see Appendix).
The technical problems encountered in  such approaches have already been studied
in details in previous references~\cite{blp92,mblw98} and will not be
described in details in this paper. We briefly discuss in Appendix the
different characteristics of the studied samples with respect to the present
model. We will now proceed to the analysis of the ED results.

\end{section}

\begin{section}{$J_{1}>0$: AF nearest neighbor interactions} 
\label{sec:3}
\subsection{The purely Heisenberg model, and phase I of the quantum model} 
\label{sec:3.1}
The AF Heisenberg point ($J_{1}=1$, $J_{2}=J_{3}=0$) 
was previously investigated by 
exact diagonalization calculations on small samples~\cite{oitmaa78},
Monte Carlo simulations~\cite{reger89},
series expansions around the Ising limit~\cite{oitmaa92},
spin-wave theory up to second order~\cite{weihong91},
and Schwinger-boson mean field theory~\cite{mattsson94}.
All concluded that the quantum system exhibits N{\'e}el LRO but with
a large reduction of the order parameter due to quantum fluctuations
(the Monte Carlo result, rather close to the spin-wave value,
is $m^{\dag}= 0.44 \pm 0.06$). 
Our ED results for sizes up to $N=32$ are consistent with this conclusion.
The approach is however different. 
In antiferromagnets with linear Goldstone modes, 
the scaling law for $m^{\dag}$ is an expansion in $1/N^{1/2}$
~\cite{fisher89,hasen93}.
For the sizes encountered in ED, the asymptotic $1/N^{1/2}$ law
is never reached \cite{blp92} and the extrapolation of the order parameter to
the thermodynamic limit remains uncertain. A qualitative idea of the finite size
scaling  of $m^{\dag}$  can be obtained from the LSW results: they show that the
asymptotic $1/N^{1/2}$ regime  cannot be expected for sizes smaller than $N \sim
400$ (see Fig. \ref{fig-op-af}).

Nevertheless confirmation of N{\'e}el LRO can be  obtained thanks
to characteristic features 
of the spectrum itself which have more favorable scaling behaviors:

\begin{itemize}
\item For a given sample size, the lowest eigenlevels in each 
sector of the total spin $S$ evolve as $E_{0}(S,N)\propto S(S+1)$
up to $S\sim\sqrt{N}$, as shown in Fig.\ref{fig-heis-tower}. They are
the eigenlevels associated with the collective dynamics of the
order parameter, the so-called Quasi Degenerate Joint States
(QDJS)\cite{blp92},    which can be described by the effective 
Hamiltonian:
\begin{equation}
H_{coll. dyn.}= \frac {1}{2 }\Delta (N)\; {\bf S}^{\,2}
\label{eq-Hcoll}
\end{equation}
where ${\Delta} (N)$ is the finite-size difference in total
energy between the absolute
ground-state and the first triplet excitation.

\item The number of QDJS and their symmetries are those expected 
for the projections of the classical N\'eel order on the irreducible representations
(IR) of $SU(2)\otimes {\cal G}$ (${\cal G}$:lattice symmetry group):
There is just one state for each $S$ value
since there is just one way to couple two spins of magnitude $N/4$
in a total spin $S$.
These states are invariant under lattice translations
(their wave-vector is ${\bf k}=0$),
under $2\pi/3$ rotations around the center of an hexagon,
they are even (odd) under inversion with respect
to the center of an hexagon for  even  $S$ and $N=4p$
(respectively odd $S$ and $N=4p+2$).  

\begin{figure}
	\begin{center}
	\resizebox{8cm}{!}{
	\includegraphics{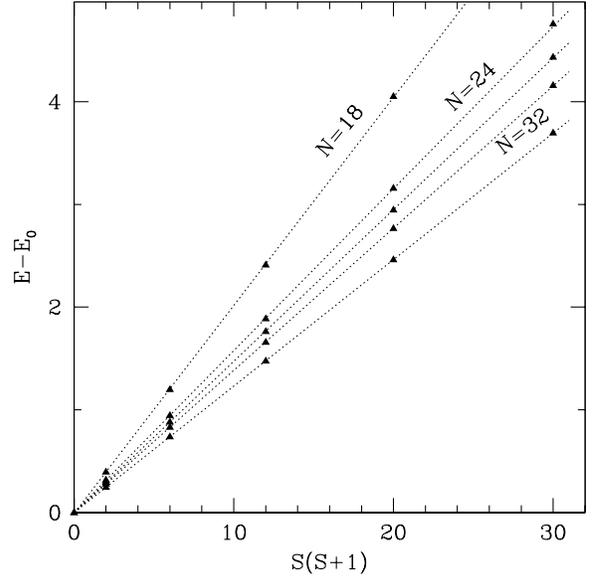}}
	\end{center}

    \caption[99]{ AF Heisenberg model, scaling of the QDJS with $S$
 and  $N$ for $N=18,24,26,28,32$.}
    \label{fig-heis-tower}
\end{figure}

\begin{figure}
	\begin{center}
	\resizebox{8cm}{!}{
	\includegraphics{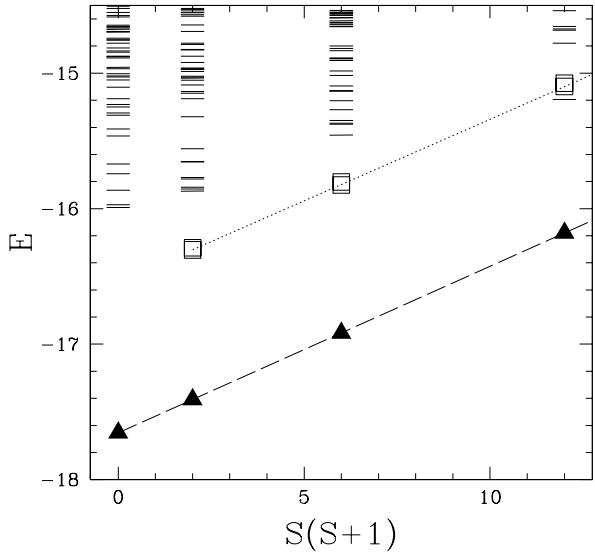}}
	\end{center}

    \caption[99]{ Low energy part of the  AF Heisenberg spectrum  for $N=32$:
eigenenergies are plotted versus eigenvalues of ${\bf S}^2$.
Full triangles represent QDJS; empty squares describe the softest magnon.
The dashed-line and the dotted-line are guides to the eye for the QDJS 
and the softest magnon respectively.
 }              
    \label{fig-heis-spec}
\end{figure} 
\item 
In the thermodynamic limit, the QDJS collapse on the singlet ground-state as $1/N$ (Fig. \ref{fig-pure-gspin}b).
The QDJS  remain distinct from the softest magnon excitation which collapse to the ground-state as $1/N^{1/2}$.
The QDJS and the softest magnon are shown for the $N=32$ sample in Fig.\ref{fig-heis-spec}.
\item
The asymptotic $\sim 1/N$ behavior of $\Delta (N)$ is not yet reached for our
largest sizes as seen in Fig.\ref{fig-pure-gspin}b.
The next order term in the $1/N^{1/2}$ expansion reads
~\cite{fisher89,hasen93}:
\begin{equation}
\Delta (N)= \frac{1}{4 \chi N} (1 - \beta\frac{c}{\rho \sqrt{N}}) 
+ {\cal O} (\frac{1}{N^2})
\label{eq-slope_tower}
\end{equation}
where  $\chi$ is the spin susceptibility, $c$ is the spin-wave velocity, $\rho$ the spin stiffness
and $\beta$ is a number of order one.
A fit of the spin-gaps to this scaling law is shown 
in Fig.\ref{fig-pure-gspin}b.
The importance of the term in $1/{N}^{3/2}$ 
is not unexpected since this term is $\propto c/\rho$ 
and quantum fluctuations which reduce $\rho$  with respect to
its classical
value are strong 
(as already shown by the reduction of the order parameter).

\begin{figure}
	\begin{center}
	\resizebox{8cm}{!}{
	\includegraphics{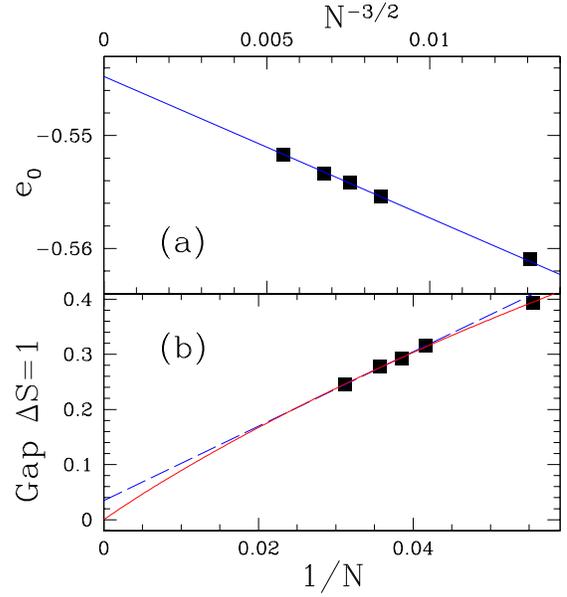}}
	\end{center}
    \caption[99]{ AF Heisenberg model, (a) energy per site $e_0$:
the line is a fit to the leading term of eq.11.
(b) spin-gap:
the full line is a fit to eq.10,
the  dashed line if a linear fit in $1/N$.
}
    \label{fig-pure-gspin}
\end{figure}

\item  The energy per site $e_{0}(N)=E_{0}(0,N)/N$ of the ground-state
scales as: 
\begin{equation}
e_{0}(N) = 
e_{\infty} - \frac{\alpha'}{N^{3/2}} (1 - \beta' \frac{c}{\rho\sqrt{N}}) 
+ {\cal O} (\frac{1}{N^{5/2}}).
\label{eq-e_N}
\end{equation}
 for a N\'eel order. 
Fig.\ref{fig-pure-gspin}a shows that the leading term
of order ${\cal O}(1/N^{3/2})$  is enough to describe the size effects
in this range. 
\end{itemize} 

A rapid analysis of the quantum phase diagram in region I does not reveal new
phases, but both LSW calculations and ED confirm that a weak antiferromagnetic
second or ferromagnetic third neighbor coupling are 
sufficient to kill LRO: the boundary between
phase I and phases III and V is shifted upwards by quantum effects.

\subsection{Region IV} 
\label{sec:3.2}
In region IV the classical model  presents a degenerate manifold of 
four-sublattice ordered ground-states. 
The finite-size spectra clearly show that this degeneracy is lifted by 
quantum fluctuations  which  favor a collinear
two-sublattice order (see Fig.\ref{fig-af_j3_200_spec}):
the low lying levels of these spectra, below the magnons excitations,
 exhibit a large family $\{ ^4\tilde E\}$ of QDJS
  associated to four-sublattice solutions.
At the bottom of this family there appears 
a line of eigenlevels with definite symmetries: these levels 
 constitute the family $\{ ^2\tilde E\}$
of QDJS states associated to a collinear symmetry breaking.

\begin{figure}
	\begin{center}
	\resizebox{8cm}{!}{
	\includegraphics{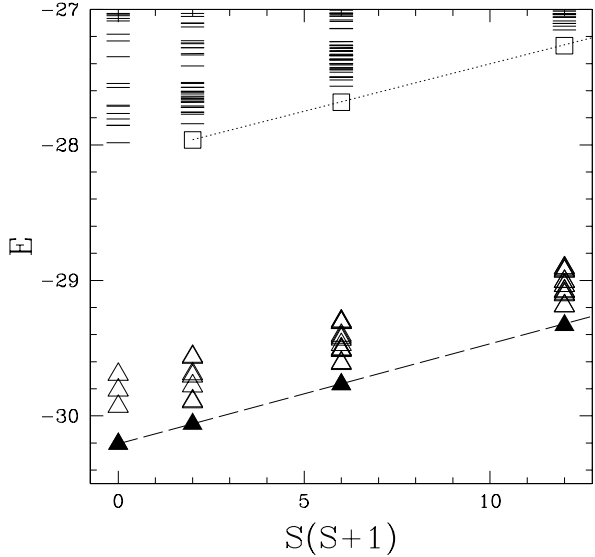}}
	\end{center}
\caption[99]{  Low energy spectrum for $J_1=1,J_2=0,J_3=-2$ and $N=32$. Full triangles represent states of the family $\{ ^2\tilde E\}$,
empty triangles represent states belonging to $\{ ^4\tilde E\}$ and not to $\{ ^2\tilde E\}$, and empty squares represent the softest magnon. All theses states have the symmetries predicted in Tables 2 and 3.
}
\label{fig-af_j3_200_spec}
\end{figure}

The situation, seen here, is very similar to the one previously studied
for the $J_1-J_2$ antiferromagnet on  the triangular lattice~\cite{lblp95a}.
The expected number of states $^4N_S$, $^2N_S$
in  $\{ ^4\tilde E\}$ and $\{ ^2\tilde E\}$ 
and their space symmetries are easily determined
for each value of the total spin $S$.
The eigenstates of $\{ ^4\tilde E\}$ can be labelled by the 
five irreducible representations $\Gamma_i$ ($i=1,5$)
of $S_4$ (permutation group of four elements).
The mapping between the space group operations on 
the four-sublattice solutions and permutations of $S_4$ is
described in Table \ref{Table-1}, together with the character Table of $S_4$.
The four-sublattice order is invariant in two-fold rotations
($\cal{R_{\pi}}$): thus the eigenstates of  $\{ ^4\tilde E\}$ belong to
the trivial representation of $C_2$. 
Since it is also invariant under a two-step translation of the Bravais 
lattice they have either a wave-vector ${\bf k=0}$ or a wave-vector 
$ {\bf K}_{i} $. 
$\Gamma_1,\Gamma_2$ and $\Gamma_3$ belong to the
${\bf k=0}$ subspace, whereas $\Gamma_4$ and
$\Gamma_5$ belong to the subspace  $\{  {\bf K}_i \}$. 
$\Gamma_1$ and $\Gamma_2$ are invariant under 
the three-fold rotations ${\cal R}_{2\pi/3}$ of $C_3$,
whereas $\Gamma_3$ is associated with the two-dimensional 
representation of $C_{3v}$.
The number of replicas of $\Gamma_i$ that should appear
for each $S$ value can be computed as in ref~\cite{lblp95a}. The result are shown in Table  \ref{Table-2} for the N=32 sample.
Analysis of the two-sublattice order can be done similarly:
the collinear solution has a three-fold degeneracy, 
the set of eigenstates $\{ ^2\tilde E\}$ maps on $Z_3$. 
It is characterized by the IR $\Gamma_1$, $\Gamma_3$ and $\Gamma_4$. The number of repliquas are shown in Table  \ref{Table-3} for the N=32 sample.  

\begin{figure}
	\begin{center}
	\resizebox{8cm}{!}{
	\includegraphics{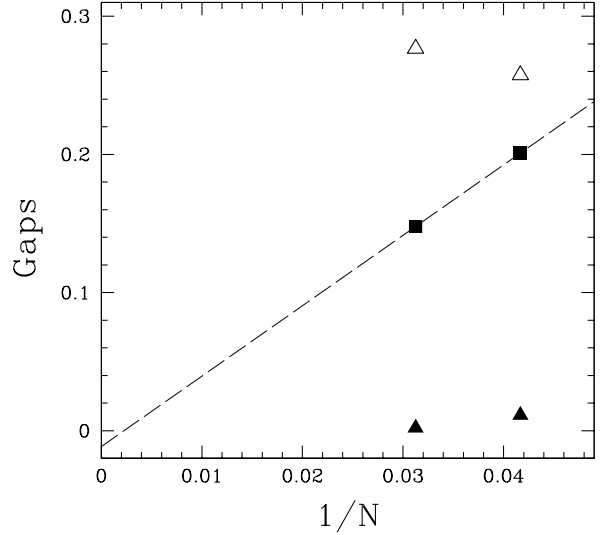}}
	\end{center}

\caption[99]{ $J_1=1,J_2=0,J_3=-2$, energy gaps measured from the
 absolute ground-state versus $1/N$ for $N=24,32$.
Full squares connected by the dashed line: gap to 
the lowest energy state in the triplet sector (it belongs to $\{ ^2\tilde E\}$).
Full triangles: gap to the $2^{nd}$ singlet state of symmetry $\Gamma_3$ (it belongs to $\{ ^2\tilde E\}$).
Open triangles: gap to the $3^{rd}$ singlet state of symmetry $\Gamma_3$
(this state belongs to $\{ ^4\tilde E\}$ and not to $\{ ^2\tilde E\}$).
}
\label{fig-af_j3_200_gaps}
\end{figure}

The ``order out of disorder'' phenomenon~\cite{villain77}
is clearly seen for $J_1=1,J_2=0,J_3=-2$.
In Fig.\ref{fig-af_j3_200_spec} 
we show the lower part of the $N=32$ spectrum at this point.
The lowest eigenstates in each $S$ sector are the states of $\{ ^2\tilde E\}$, describing collinear order.
Further support  to  this assumption 
 is given by the finite size effects of the energy gaps.
As shown in Fig.\ref{fig-af_j3_200_gaps}, a plot of the spin-gap of the 
$N=24,32$ samples  versus $1/N$ is consistent with a vanishing value 
for $N\to\infty$. On the other hand
the gap between the two states $\Gamma_1$ and $\Gamma_3$ 
of $\{ ^2\tilde E\}$ of the $S=0$ sector
tends to close when the size goes to infinity,
 whereas the gaps between the levels of $\{ ^2\tilde E\}$ and 
the other levels of $\{ ^4\tilde E\}$ increase with $N$.

We  have investigated the scaling behavior of the spectra at some other points of
region  IV not too close from the classical boundaries and found essentially
 the same behavior and a selection of collinear LRO by quantum fluctuations.

Closer to the boundary 
between region IV and V, the separation between the $\{ ^4\tilde E\}$
states and the magnons states decreases. This is an indication of
 a softening of the magnons and  the neighborhood of
a 2nd order phase transition towards another phase.
The behavior of the spin-gaps at $J_1=1,J_2=0.5,J_3=-0.5$ and
$J_1=1,J_2=0,J_3=-1$, similar to Fig.\ref{fig-af_j3_200_gaps}, nevertheless
  indicates that these points of the quantum phase diagram are still in the
collinear phase IV. 

Various studies of the spectra of the $N=18,26,28$ samples under suitable
boundary conditions confirm these results for the quantum phase IV, and indicate
that the quantum boundary between phase IV and V is probably  slightly shifted down
relatively to the classical boundary shown in Fig.2.

Results of ED calculations (not shown) in region II for $J_1<0$ 
(which has the same classical manifold of degenerate 
ground-states as in IV for $J_1>0$)
suggest a similar selection of the collinear solution there too.
In conclusion up to a slight motion of the boundaries , the semi classical
behavior in regions I, II, IV and VI, is not qualitatively affected by the
strong quantum fluctuations of the spins 1/2.

\subsection{Quantum phases between I and IV} 
\label{sec:3.3}
The intermediate phases  between the two collinear N{\'e}el phases 
cover region V and part of region III.
In this part of the quantum phase diagram, $SU(2)$ symmetry is
unbroken, and there is a gap to triplet excitations: these phases only
support short range order in the spin-spin correlations 
(see Fig.\ref{fig-cor-spin}).
\begin{figure}
\begin{center}
\resizebox{8.5cm}{!}{
\includegraphics{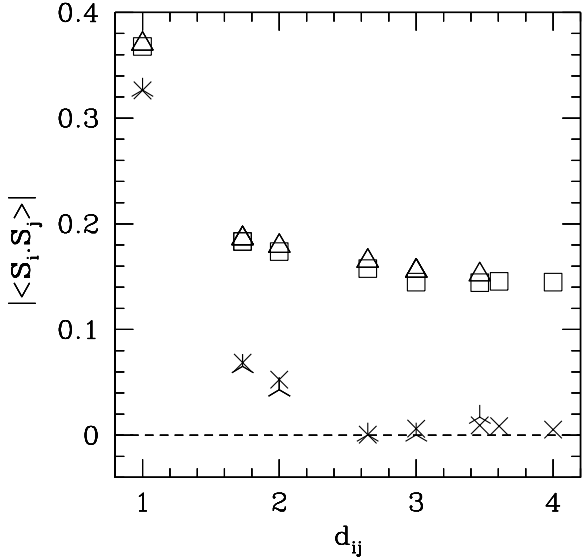}}
\end{center}
    \caption[99]{Spin-spin correlations as a function of distance 
for the pure Heisenberg model on N=24 (triangles) and N=32 samples (squares) and for $J_1=1, J_2=0.3$
on N=24 (three-legged star) and N=32 samples (four-legged star).
}
    \label{fig-cor-spin}
\end{figure} 
Our LSW and ED calculations indicate that this quantum region likely extends 
in regions I and IV
\footnote{LSW calculations predict non vanishing
order parameters for the classical spiral solutions inside III 
away from the boundaries but a vanishing order parameter in the whole
region V.}.
In this work we study region V, region III  close to I, and the 
transition line III-V with ED calculations 
using TBC on $N=18,24$ samples (see Appendix) and PBC on 
the $N=24$ and $N=32$ samples.
A thorough search of the ED spectra, sweeping the twist angles
at the sample boundaries, did not yield evidence of incommensurate
helical LRO, neither with the wave-vectors of the classical solutions
nor at other wave-vectors.
In all cases no tower of QDJS was found.
The ED results corroborate the conclusion of LSW calculations
that the classical spiral solutions
are destabilized by quantum fluctuations. This seems a rather
general statement in systems where the quantum fluctuations are
strong enough\cite{moreo90,read-sachdev91}.
\begin{figure}
	\resizebox{8cm}{!}{
	\includegraphics{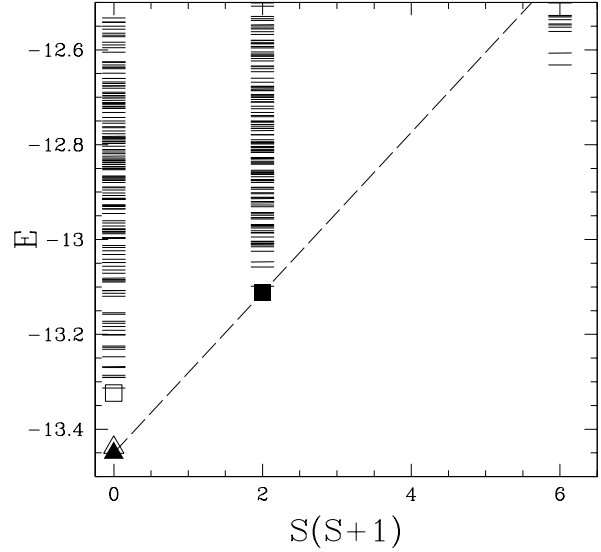}}

\caption[99]{ Low energy spectrum for $J_1=1,J_2=0.4,J_3=0$ and
$N=32$. Full triangle: ground state; empty triangle: first singlet
excited state (these two states have a wave vector ${\bf k}=0$); empty
square: second singlet excited state; full square: first triplet state.}
\label{fig-11}
\end{figure}

Is this quantum phase a quantum  disordered one? 
To answer this question
we performed extended ED calculations on $N=24,32$ samples
on different points of the transition line III-V  where the classical model
has an infinite set of spiral ground-states, and the LSW calculation
diverges. Along this line, we found evidence of  two different phases 
both with a gap.
 
Let us begin by the phase around  $J_2=0.4$: 
this point is very close to the point where the energy versus $J_2$ is the largest
and may be considered as a point of maximum frustration.
The spectrum of the $N=32$ sample is shown in Fig.\ref{fig-11}.
This spectrum 
differs from the spectra of the collinear ordered system in IV:
\begin{itemize}
\item the lowest states  are not IR of $\{ ^2\tilde E\}$
\end{itemize}
and features associated with N{\'e}el LRO are missing:
\begin{itemize}
\item The lowest eigen-energies for each $S$ value  do not 
increase as $S(S+1)$ with $S$
\item The lowest states in each $S$ sector are not separated from the others
as the QDJS are separated from the magnons, instead
there is a dense continuum of states  in each $S$ sector
except the $S=0$ one.
\item Furthermore a plot of the spin-gap of the $N=24,32$ samples versus $1/N$,
displayed in Fig.\ref{fig-12}, shows that the scaling law characteristic 
of a N{\'e}el ordered system is not obeyed and 
indicates a large spin-gap $\approx 0.2$ for $N\to\infty$.
\end{itemize}

\begin{figure}
	\begin{center}
	\resizebox{8cm}{!}{
	\includegraphics{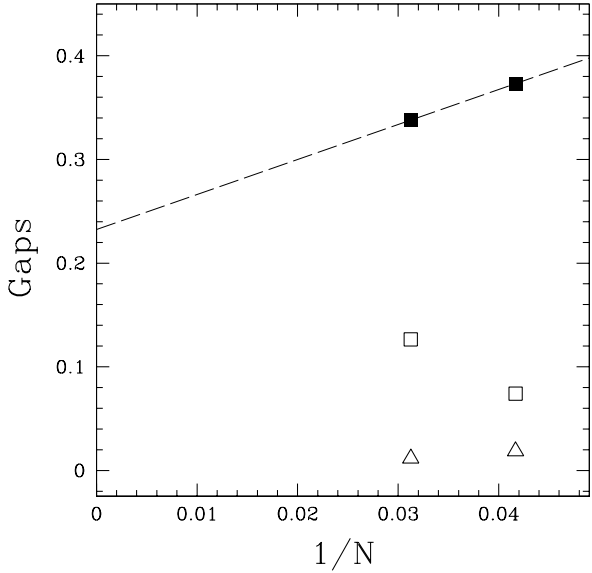}}
	\end{center}
\caption[99]{ $J_1=1,J_2=0.4$, energy gaps measured from the absolute 
ground-state versus $1/N$ for $N=24,32$.
Full squares: spin gap, i.e. gap to the first triplet excitation;
open triangles:
gap to the first singlet excitation 
($ {\bf k}=0$, IR $\Gamma _3$); open squares:
gap to the second singlet excitation.}
\label{fig-12}
\end{figure}   

Most likely however the system is not fully disordered but exhibits dimer LRO
(see Fig.\ref{fig-13}).
The dimer operator on a pair of sites $\left(i,j\right)$ is 
$d_{i,j}=\left(1-P_{i,j}\right)/2$
where $P_{i,j}=2({\bf S}_i.{\bf S}_j +1/4)$ is the spin permutation operator.
This projector is greater
(less) than $0.25$ when the spin-spin correlation is negative
(positive), equal to 1 on a singlet and to 0 on a triplet.
For $J_1=1,\ J_2=0.4$, on the $N=32$ sample, the first neighbor correlation is
$<d_{k,l}>=0.4899$.
On the  symmetry breaking Spin-Peierls state 
(pure product of ordered dimers),
the average value of the dimer operator is 1 on the bonds where there
is a dimer, and $1/4$ on the other bonds.
As the exact eigenstate does not break $C_3$ symmetry the number $0.4899$ 
 should be compared with the result obtained on the
symmetric superposition of the three Spin-Peierls states aligned
 along the three main
directions of the lattice: in this symmetrized Spin-Peierls state
this correlation is $d^{\Psi_{sym}}_{k,l}=0.5$.
The average value of the dimer operator in the exact ground-state is thus
very close to the Spin-Peierls value.
 
The dimer-dimer correlation  between a reference bond $(i,j)$
and the bond $(k,l)$ is 
$D_{(i,j),(k,l)}=<d_{i,j}d_{k,l}>-<d_{i,j}><d_{k,l}>$.
As in ref~\cite{mblw98}, we normalized $D_{(i,j),(k,l)}$ by its maximum value 
which is achieved when the two bonds are completely correlated.
We  thus measured dimer correlations by
\begin{eqnarray}
p_{(i,j),(k,l)} &=& 
 \frac{D_{(i,j),(k,l)}}{<d_{k,l}>-<d_{i,j}><d_{k,l}>} \\ \nonumber
&=& \frac{<d_{i,j}d_{k,l}>-<d_{i,j}><d_{k,l}>}
         {<d_{i,j}>\left(1-<d_{k,l}>\right)} 
\label{eq-dimer}
\end{eqnarray}
If $p_{(i,j),(k,l)}=1$ the presence of a dimer on bond $\left(i,j\right)$ 
implies the existence of a dimer on  
bond  $\left(k,l\right)$; if $p_{(i,j),(k,l)}=0$ there is an absence of 
correlations between singlets on bonds $(i,j))$ and
$(k,l)$. If $p_{(i,j),(k,l)}$ is negative,
 a singlet on bond $(i,j)$ induces a tendency towards  ferromagnetic
correlation on bond $(k,l)$.

The correlation pattern 
for dimer on first neighbor bonds is displayed in  Fig.\ref{fig-13}.
\begin{figure}
	\begin{center}
	\resizebox{8cm}{!}{
	\includegraphics{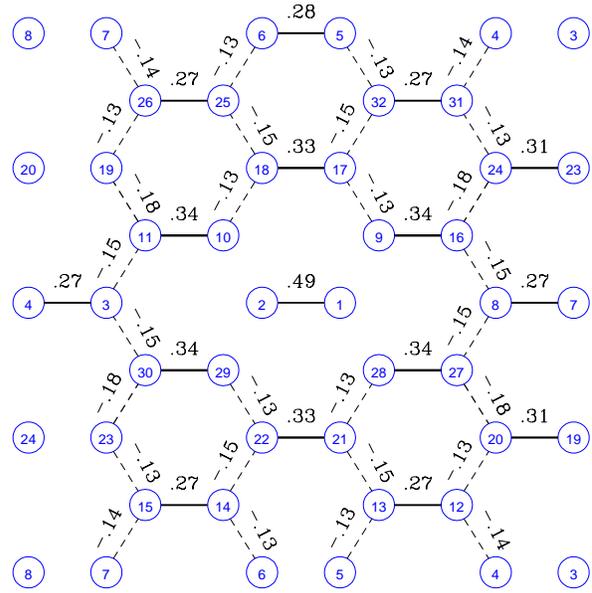}}
	\end{center}

\caption[99]{ $J_1=1,J_2=0.4$,  singlet-singlet correlations $p_{(1,2),(k,l)}$
between the reference bond $\left(1,2\right)$ and bonds $\left(k,l\right)$
in the ground-state of the $N=32$ sample.
The numbers above bonds $\left(k,l\right)$ are the values of $p_{(1,2),(k,l)}$
truncated to the two first significant digits.
Full (dashed) lines  indicate positive (negative) values of $p_{(1,2),(k,l)}$, 
The width of the lines is proportional to the magnitude of $|p_{(1,2),(k,l)}|$.
The number above  the bond $\left(1,2\right)$ is $<d_{1,2}> $ (see text).
}
\label{fig-13}
\end{figure}
The calcul of dimer-dimer correlations on the state $\Psi^{S.P}_{sym}$ gives 
$p_{(i,j),(k,l)}=+0.5 $ if (i,j) and (k,l) are
parallel and   $p_{(i,j),(k,l)}=-0.25$ if (i,j) and (k,l) are
non parallel bonds.
The exact dimer-dimer correlations are  not too far from these values and
decay very slowly with distance.
This is in favor of a columnar LRO of dimers with a
$C_3$ symmetry breaking, previously proposed by
Einarsson and coll.~\cite{ej91}.

Moreover the degeneracy of the ground-state for $N\to\infty$ points to the same
conclusion:  Fig.\ref{fig-12} indicates that
the gap between the ($\Gamma_1$ and $\Gamma_3$, $S=0$) lowest states,
which are both ${\bf k=0}$ states, 
closes for $N~\to~\infty$, while the gap between these states
and the upper levels increases with the size.
$\Gamma_1$ is non degenerate and $\Gamma_3$ 
twice degenerate:  this allows the building of the 
three columnar dimer patterns with a $C_{3}$ symmetry breaking  and no
translation breaking. In this picture the finite size ground-state
is the symmetric combination of these three states.
This degeneracy corresponds to a true symmetry breaking with 
a local non zero order parameter (dimer LRO):
this is a Valence-Bond Crystal(VBC).

The honeycomb lattice could a priori accommodate a different kind of 
VBC with alternation of hexagons with three dimers 
and hexagons without dimers:
this pattern breaks both $C_3$ and translational symmetry. In their
large N approach, Read and Sachdev~\cite{rs90} found that this structure
might be the ground-state. In the range $J_2 \approx0.3-0.35$,  we find
 a short range  structure roughly reminiscent of this
arrangement.
In fact the short range dimer-dimer 
correlations are even more symmetric than
in this VBC crystal and would be more compatible with a crystal
of hexagon-plaquettes in a symmetric $S=0$ state\cite{mo01}: for example the
 correlation (1-2)(7-6) should be negative and equal
to $-0.25$ in the Read and Sachdev VBC state whereas it is equal
to $0.1$ in the pure hexagon-plaquette VBC.
In fact in our $SU(2)$ model
 all  correlations decrease noticeably with distance 
(Fig. \ref{fig-cor-spin}, \ref{fig-cor-dim-d30}) 
 and the pattern does not seem to propagate at large distances.
The ground state in this range of parameter is probably a 
RVB spin liquid.
This  conclusion is qualitatively
substantiated by the study of the energy
gaps to the ground-state: plausibly none of them goes to zero at
the thermodynamic limit, which would be consistent with the RVB
hypothesis. Unfortunately the finite-size effects are rather chaotic:
 the ground-state energy of the $N=18$ and $N=30$
samples (samples on which the Read and Sachdev dimer pattern is not
frustrated) are
larger than the ground-state energy of the $N=26$ and $N=32$ samples,
which frustrate it. This is probably
 related to the fact that the $N=18$ and
$N=30$ samples do not allow the system to take full advantage of 
the second neighbor antiferromagnetic coupling, whereas the $N=26$ and
$N=32$ samples do. But the building of  singlets
 on second neighbor bonds tends
to destroy VBC patterns and favor a RVB ground-state.
All these arguments point in favor of an RVB phase at this
coupling parameter: unfortunately the sizes that can be studied do not
allow a quantitative determination of the gaps.

\begin{figure}
	\begin{center}
	\resizebox{8cm}{!}{
	\includegraphics{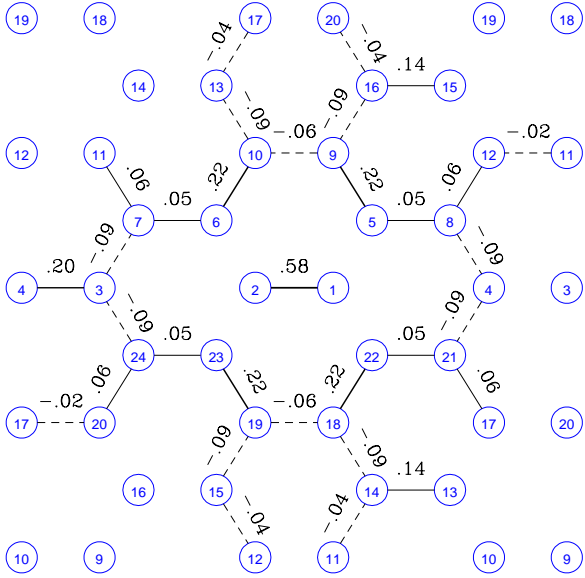}}
	\end{center}
\caption[99]{ 
 Singlet-singlet correlations for $J_1=1,J_2=0.3$ and $N=24$
 (same legend as in Fig.\ref{fig-13}) .
}
\label{fig-cor-dim-d30}
\end{figure}

The quantum AF $J_1-J_2$ model on the square lattice 
close to  the point of maximum
frustration exhibits dimer~\cite{singh99} or plaquette~\cite{capriotti00}  LRO;
the same kind of conclusion has been drawn for the $J_1-J_3$ model~\cite{ll96},
and for the MSE model on the square lattice~\cite{cgb92}.
These phases share qualitative properties with the phase identified for 
$J_2=0.4$: in each cases a collinear LRO is
destabilized by frustration giving birth through a 2nd-order phase transition
to a massive phase with dimer LRO.
Such  VBC phases appear in many  models on
bipartite lattices:  Rokhsar and
Kivelson~\cite{rk88} in the Quantum Dimer approach (QD),
Dombre and Kotliar~\cite{dk89} for the Hubbard model,
Read and Sachdev~\cite{read-sachdev91} in the $SU(N)$ approach
of the Heisenberg model found VBC phases.
These phases, as the  first one described here (for $J_2=0.4$),
 have a gap for all
excitations, a discrete degeneracy of the ground-state,
exponential decrease of the two points spin-spin correlations but 
LRO in higher correlation functions; they have  only 
confined spinons.

Up to now we only know few
spin-1/2 models exhibiting true RVB phases with a clear-cut gap:
the MSE model on
the triangular lattice\cite{mblw98} and the Quantum Dimer model on the
triangular lattice\cite{ms00}. More work is needed to know if the
excitations of these different RVB states are similar  and in
particular if they sustain deconfined spinons excitations. 

\end{section}

\begin{section}{$J_{1}<0$: ferromagnetic nearest neighbor interactions} 
\label{sec:4}

As already underlined above in Sect.~\ref{sec:3.2} 
the classical collinear phases (F or AF) 
observed for large $J_{2}$ and $J_{3}$ are likely to survive to quantum
fluctuations. We thus focus our study on 
the region of  maximum frustration,
$0 < {| J_2|},{| J_3|}<0.5$, corresponding to region V and part of
region III
of Fig.\ref{fig-class-phas-b}. In this situation 
LSW calculations predict a non vanishing order parameter of
the spiral solutions for values of $J_2$ and $J_3$ not too close
to the transition lines.
However extensive ED calculations performed on the $N=18,24$ samples
with twisted boundary conditions 
do not yield any evidence of spiral LRO.

\begin{figure}
	\begin{center}
	\resizebox{8cm}{!}{
	\includegraphics{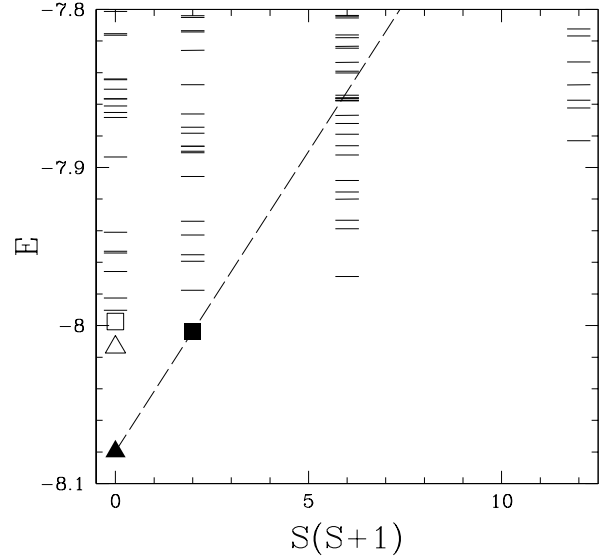}}
	\end{center}

\caption[99]{ Low energy spectrum for $J_1=-1,J_2=0.25,J_3=0$ and $N=32$. Full triangle: ground state; empty triangle: first singlet
excited state (these two states have a wave vector ${\bf k}=0$); empty
square: second singlet excited state; full square: first triplet state.
}
\label{fig-14}
\end{figure}

We thus studied samples up to $N=32$ spins
to investigate the nature of the ground-state for a few sets of parameters.
The most extensive calculations were done at the point $J_{2}=.25$, $J_{3}=0$ 
on the transition line III-V which may be considered as highly
frustrated  
(for this  $J_{2}$  value  the ground-state energy is close to its maximum,
and in the LSW approach quantum fluctuations destroy LRO).
Strong indications that the model has a  RVB spin-liquid ground-sate,  
were found at this point: 
\begin{itemize}
\item
The spectra do not exhibit a tower of QDJS as shown in Fig.\ref{fig-14}
for $N=32$,
and  $E_{0}(S)$ clearly does not evolve as $S(S+1)$ with $S$.
\item A plot of the spin-gap versus $1/N$, shown in Fig.\ref{fig-gap-f-d25}, indicates
that the spin-gap  is  small but finite  when $N\to \infty$
\footnote{ In view of Fig.\ref{fig-gap-f-d25}, 
one may object to our extrapolation to $N\to \infty$
on two numerical samples. In fact our conclusion is supported by examination
of both the gap and the energy per bond of the samples with sizes 18, 24,
28, 32 with various boundary conditions (available on request at:
fouet@lptl.jussieu.fr). This study shows that the variations of these
quantities with the size is very small and mainly due to the frustration of
the short range antiferromagnetic order between third neighbors due the
boundary conditions (see below) and {\it not} to the cut-off in
the low-energy long wave-length quantum fluctuations. Notice that the energy
per bond on the two non frustrating sizes 24 and 32 does not increase with
the system size but decreases by a very small amount ($\sim 10^{-3}$). We thus
conclude that we are in the cross-over regime for both sizes 24 and 32 and
the extrapolation of the spin gap in Fig.\ref{fig-gap-f-d25} is reasonable.
}.
\end{itemize}

\begin{figure}
	\begin{center}
	\resizebox{8cm}{!}{
	\includegraphics{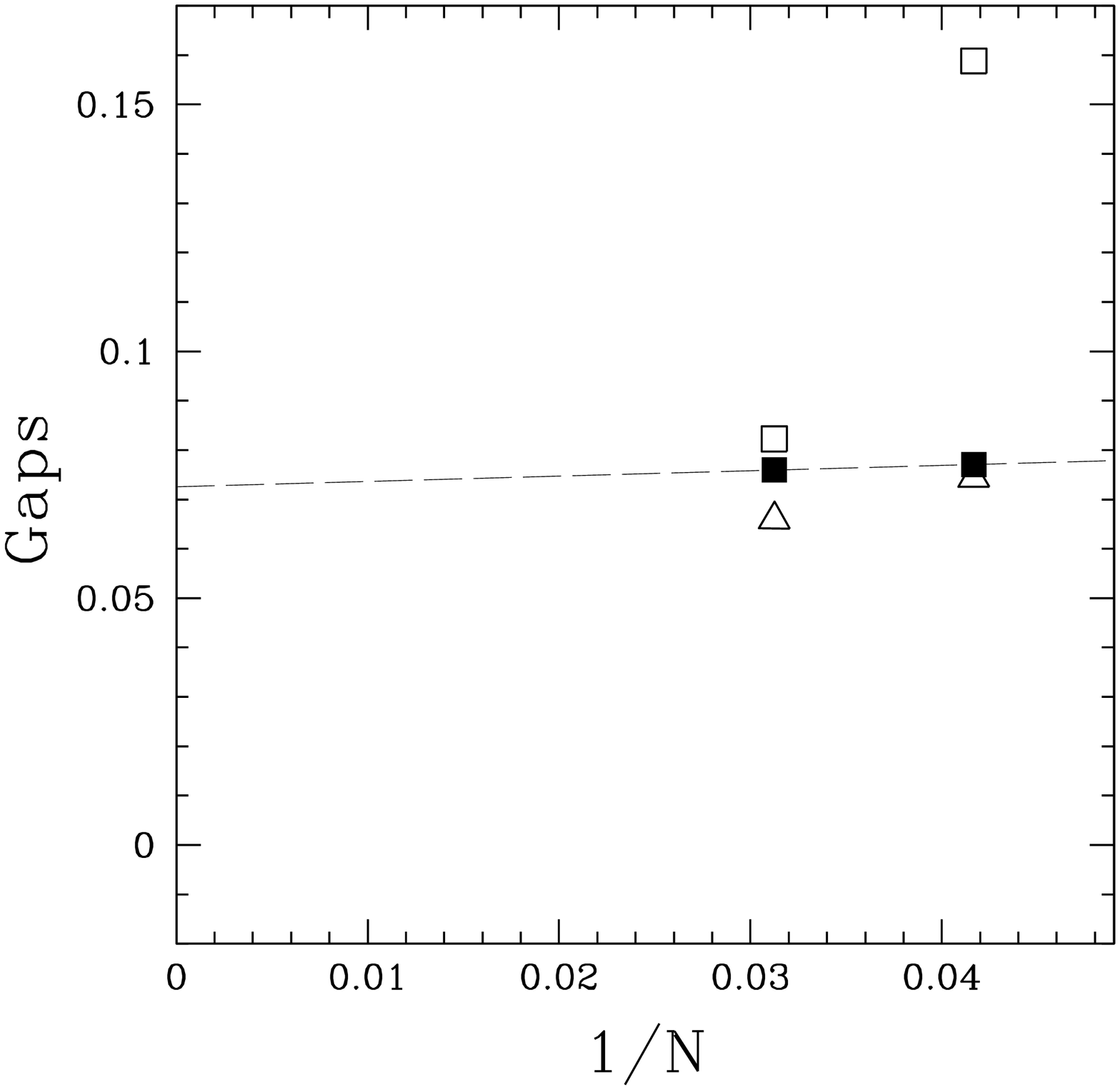}}
	\end{center}
\caption[99]{ $J_1=-1,J_2=0.25$,  energy gaps measured from the absolute ground-state vs $1/N$ for $N=24,32$.
 Black squares show the spin-gap. Open triangles (squares)
 the gap to the first (second) excitation in the
singlet sector.  }
\label{fig-gap-f-d25}
\end{figure}

But contrary to the case with positive $J_{1}$ and $J_2=0.4$:
\begin{itemize}
\item The correlations display a strong short range order but plausibly no LRO.
The short range pattern is original: the first neighbor spin-spin 
correlation is ferromagnetic ({$ <{\bf S}_i.{\bf S}_j> = 0.10$}),
the second (third) neighbor spin-spin correlations are  
antiferromagnetic 
(
\mbox{$ <{\bf S}_i.{\bf S}_j>_{n.n.}$}$ =-0.13 $, 
\mbox{$ <{\bf S}_i.{\bf S}_j>_{n.n.n.}$}$ =-.25 $
),
but no long range pattern does emerge from this
picture. The dimer-dimer correlations equally show a strong short range
pattern and apparently no LRO.
Fig. \ref{fig-cor-dim-f-d25} represents first-neighbor dimer-dimer
correlations: they are much weaker than in the AF case (remark that
 triplet-triplet correlations are equal to singlet-singlet ones and
 compare with Fig. \ref{fig-13}).
 Fig. \ref{fig-cor-dim-f-d25-2v} displays second-neighbor
 dimer-dimer correlations which also decrease with distance.
 The  third-neighbor dimer-dimer correlations decrease even quicker.
  The strength of the short range correlations
explains the finite size results on small frustrating sizes
 (see footnote $^4$).  

\begin{figure}
	\begin{center}
	\resizebox{8cm}{!}{
	\includegraphics{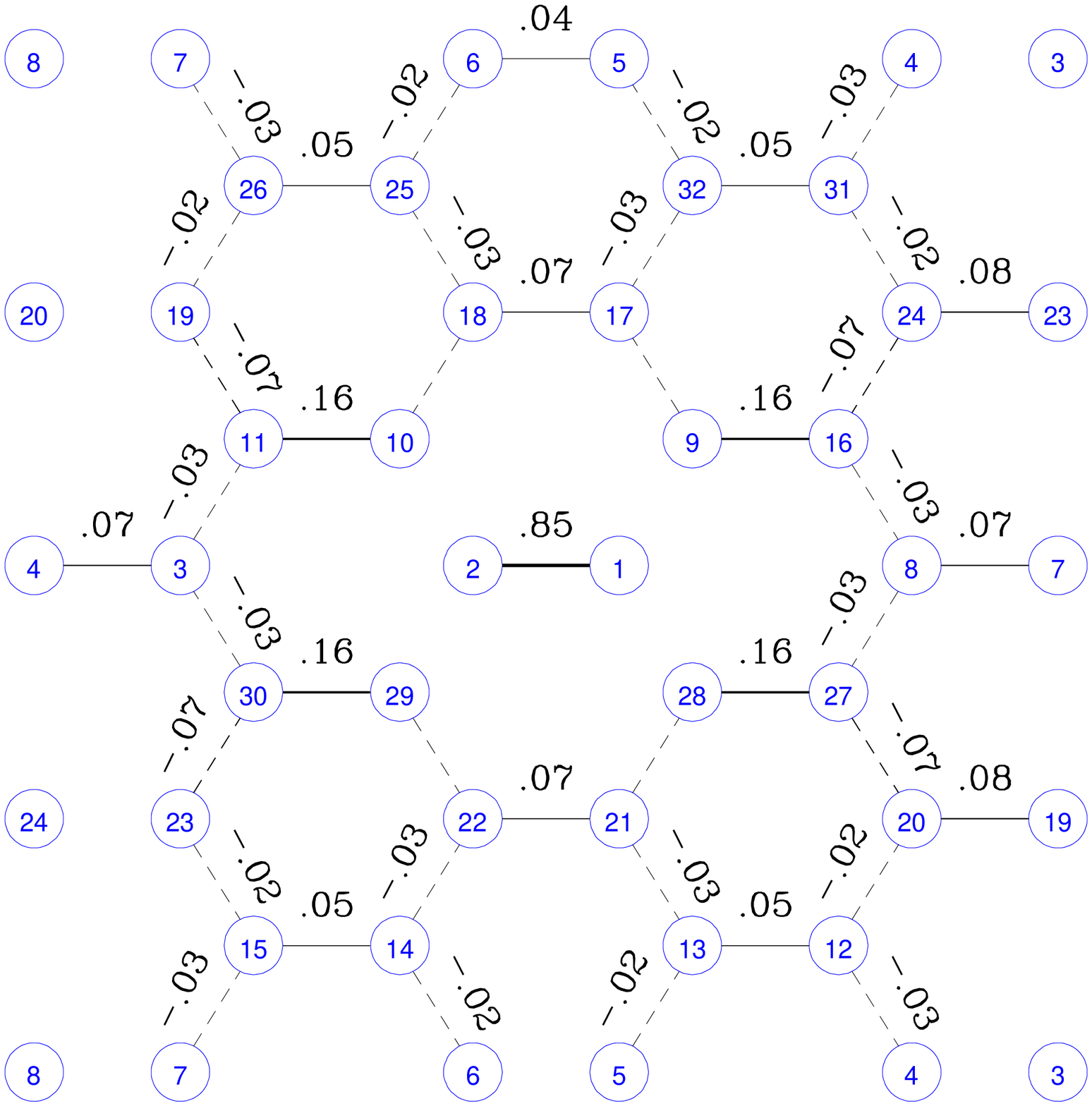}}
	\end{center}

\caption[99]{ $J_1=-1,J_2=0.25$, first neighbor triplet-triplet 
correlations on the $N=32$ sample. The spin-spin correlation on the
 reference bond $(1,2)$ is ferromagnetic: the number above this
bond measures the
projection of the two-spin state of the exact ground-state on the
pure triplet state.}
\label{fig-cor-dim-f-d25}
\end{figure}
\begin{figure}
	\begin{center}
	\resizebox{8cm}{!}{
	\includegraphics*[1.1cm,8.1cm][18.2cm,22cm]{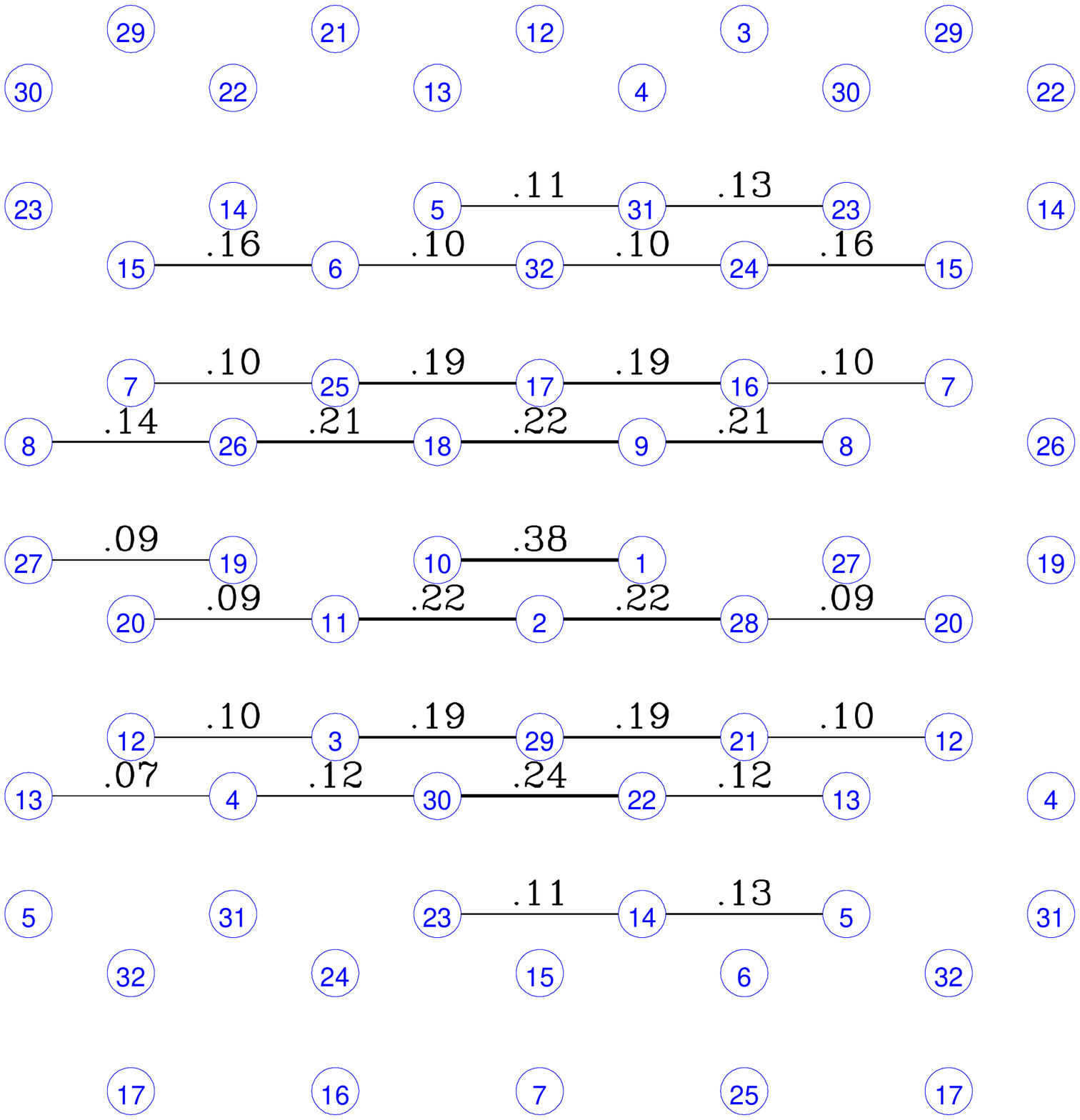}}
	\end{center}
\caption[99]{ $J_1=-1,J_2=0.25$, second neighbor  singlet-singlet
correlations on the $N=32$ sample.
The reference bond is $(1,10)$: the spin-spin correlation on this
bond is antiferromagnetic, the number above the bond measures the
projection of the two spin-state of the exact ground-state on a pure
singlet i.e.: $<d_{1,10}>$; 
 the value of this observable in a symmetrized wave function of
products of second-neighbor singlets is 0.375.
Non-parallel  singlet-singlet
 correlations have been omitted for clarity,
all of them are very small and negative.}
\label{fig-cor-dim-f-d25-2v}
\end{figure}
\item The ground-state is probably unique in the thermodynamic limit. 
The first two singlet excitations are shown together with the first triplet
excitation in Fig. \ref{fig-gap-f-d25}.
The spin-gap and  the ground-state energy per spin
display a very small sensitivity to the size for $N=24,32$.
The gap to the third
excitation seems more sensitive to the size but this might due to the fact that
the different sizes do not accommodate the same wave vectors.
In view of the results it seems probable that the singlet excitations will not
collapse to the absolute ground-state in the thermodynamic limit.

\end{itemize}

 We thus conjecture that
the quantum ground-state of this system does not break any symmetries of the
Hamiltonian or of the lattice: it is a { \it quantum spin-liquid }
 where all excitations are gapped.
Results obtained for the $N=24$ and $32$ samples for $J_{2}=.5$
also indicate a finite spin-gap when $N\to \infty$.
This  suggest that there is a spin-liquid phase 
in a  finite range of parameters.

Such a quantum massive phase, without LRO, is
highly reminiscent of the spin-liquid phase found in the
MSE model on the triangular lattice~\cite{mblw98}. Curiously
enough it appears in the two cases in the vicinity of a ferromagnetic
phase destabilized by antiferromagnetic couplings.
There is a difference in the degeneracy of the ground-state in the two
cases: whereas the ground-state on the honeycomb lattice is unique, it
has a 4-fold degeneracy on the triangular lattice. This is easily
understood as the honeycomb lattice is not a Bravais lattice and has
two spin-1/2 per unit cell. Thus the uniqueness of the ground-state in this
latter case does not contradict the
Lieb-Schultz-Matthis-Affleck conjecture \cite{affleck88}, or the
topological approach of Read and Chakraborty\cite{rc89}.

\end{section}
\begin{section}{Conclusions and conjectures}
\label{sec:5}

This study of the spin-1/2 $J_{1}-J_{2}-J_{3}$ model on the honeycomb lattice
has brought the following new results:
\begin{itemize}
\item For small frustrations $J_{2}/J_{1}$ or  $J_{3}/J_{1}$ 
less than $\sim 0.15$ or larger than $1$,
the system remains essentially classical: 
when various kinds of LRO are possible,
quantum fluctuations, as well as
thermal fluctuations in the classical case, select 
the LRO with the most
symmetric order parameter amongst the various possibilities.  
\item
The classical symmetry between the phase diagram for ferromagnetic
$J_1$ and  the phase diagram for antiferromagnetic $J_1$ discussed in
Sect.~\ref{sec:2.1} 
is destroyed by quantum fluctuations. 
\item
For the largest frustrations these models exhibit gapped phases.
\item For an antiferromagnetic first neighbor coupling, a Valence Bond
Crystal phase has been clearly evidenced around  $J_{2}/J_{1}= 0.4$.
\item For an intermediate frustration $J_{2}/J_{1}= 0.3$, an RVB
spin-liquid appears
between the N\'eel ordered phase and the VBC phase.

\item For a ferromagnetic first neighbor coupling, the present
results favor the hypothesis of a RVB spin-liquid phase in a large
range of parameters. No VBC has been found in that case.

\end{itemize}

This study of the spin-1/2 $J_{1}-J_{2}-J_{3}$ model on the honeycomb lattice,
when compared to similar approaches of $SU(2)$ Hamiltonians leads us
to formulate some conjectures on the generic behavior of such  models on
different lattices.
\begin{itemize}
\item 
In 2D the pure S=1/2 Heisenberg model is N{\'e}el ordered 
on any bipartite lattices
with coordination number $\geq 3$. It is disordered on the triangular-based
kagom{\'e} lattice which has a coordination number equal to 4.
\item Non-coplanar classical ground-states are not robust against quantum
fluctuations in the isotropic models.
\item N{\'e}el order or ferromagnetism
 disappears around the points of maximum classical frustration giving
birth to phases with spin-gap and short range spin-spin correlations.
\item 
Disappearance of a ferromagnetic phase due to antiferromagnetic frustrations
leads generically to a spin-liquid phase, with short range correlations in all observables
and a gap to all excitations.
\item
Disappearance of a collinear antiferromagnetic phase might lead to
a VBC phase either directly ($J_1-J_2$ model on the square lattice),
or through an intermediate RVB phase (this study). The spin-liquid
phase observed by Santoro et al in the spin-orbital model\cite{ssgpt99}
 might be rather similar to the RVB phase described here.

For completion we might add:
\item
Disappearance of a non-collinear phase (3-sublattice N{\'e}el phase) takes place
through a phase with a spin gap but a continuum in the singlet
sector\cite{lblps97,lmsl00}.
\end{itemize}

\end{section}

Acknowledgements:
Computations were performed at The Centre de Calcul pour la Recherche de
l'Universit{\'e} Pierre et Marie Curie and at the
Institut de D{\'e}veloppement des Recherches en Informatique Scientifique
of C.N.R.S. under contract 990076 .

\begin{section}*{Appendix: Special properties of the studied samples, 
and boundary conditions} 
The ED calculations were performed as in refs~\cite{blp92,mblw98} 
on systems of $N=18,24,26,28,30,32$ sites shown in Fig.\ref{fig-systems}.
The $N=18,24,32$ samples have the full point group symmetry of the lattice,
whereas the $N=26$ sample misses axial but still has rotational $C_3$
symmetry,  the $N=28,30$ have neither.

With periodic boundary conditions (PBC), 
all the samples are of course compatible with the ${\bf Q}=0$ order
in region I. 
In region IV, however only the $N=24$ and $32$ samples have the full symmetry
of the classical order. 
The $N=28$ sample is compatible with one collinear
solution but frustrates the two others as well as the non coplanar solutions.
The other samples are frustrating but can allow a collinear order
if twisted boundary conditions (TBC) are used. 
This is the case for the $N=18$
sample if a twist of $\pi$ is applied along
${\bf t}_{1}$ or ${\bf t}_{2}$.

To search for spiral order, we used TBC and sweep the whole range
$[0, 2\pi]$ of twist angles
$\phi_{1,2}$ in the ${\bf t}_{1}$ and ${\bf t}_{2}$ directions.
These specific boundary conditions are defined as:

\begin{equation}
{\bf S}({\bf R}_i+{\bf t}_j)=e^{i\phi_jS_z({\bf R}_i)}
	{\bf S}({\bf R}_i)
	e^{-i\phi_jS_{z}({\bf R}_i)}.
\label{twbc}
\end{equation}

This allows to look for boundary conditions which would not 
frustrate helical ground-states.
This approach was found effective for the Heisenberg
model on the triangular lattice to deal with samples frustrating
the three-sublattice N{\'e}el order~\cite{blp92}.
For such samples the ground-state energy
was found to reach its minimum for the twists which release
the frustration: at that point  the spectra
recover the characteristic features of N{\'e}el order.

A VBC with the pattern of Read and Sachdev~\cite{rs90}
(considered in Sect.~\ref{sec:3.3})  
fits in the $N=30$ sample but not on a $N=32$ sample.

\begin{figure}
	\begin{center}
	\resizebox{8cm}{!}{
	\includegraphics{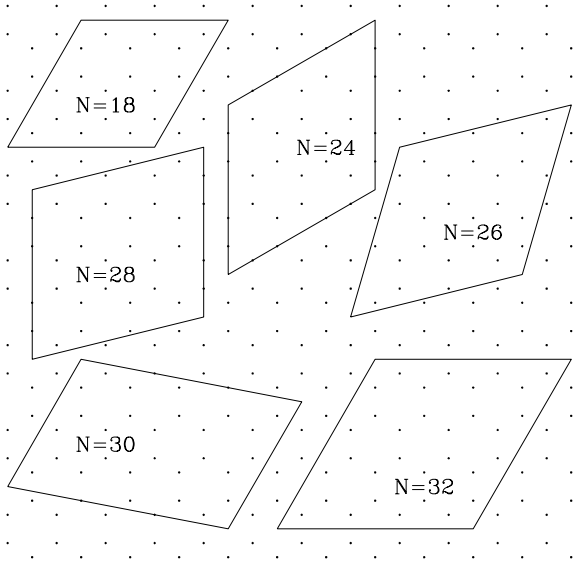}}
	\end{center}

    \caption[99]{ The $N=18,24,26,28,30,32$ samples.
}
    \label{fig-systems}
\end{figure}

\end{section}
\begin{table}
\begin{center}
\begin{tabular}{|c|c c c c c |}
\hline
$S_4$ & $ I $ & $(A,B)(C,D)$ & $(A,B,C)$ & $(A,B)$ & $(A,B,C,D)$\\
${\cal G}$ & $I$ & $t$ & ${\cal R}_{2\pi/3}$ & $\sigma$
& ${\cal R}_{2\pi/3}^{'} \sigma$ \\
  $ N_{el}$ & 1 & 3 & 8 & 6 & 6\\
  \hline
   $\Gamma_1 $ & $ 1 $ & $ 1 $ & $ 1 $ & $ 1 $ & $ 1 $ \\ 
   $\Gamma_2 $ & $ 1 $ & $ 1 $ & $ 1 $ & $-1 $ & $-1 $ \\
   $\Gamma_3 $ & $ 2 $ & $ 2 $ & $-1 $ & $ 0 $ & $ 0 $ \\
   $\Gamma_4 $ & $ 3 $ & $-1 $ & $ 0 $ & $ 1 $ & $-1 $ \\
   $\Gamma_5 $ & $ 3 $ & $-1 $ & $ 0 $ & $-1 $ & $ 1 $ \\
\hline
\end{tabular}
\end{center}
\caption[99]{Character table of the permutation group $S_4$.
First line indicates classes of permutations.
Second line gives an element of the space symmetry class
corresponding to the class of permutation. These space symmetries are: $t$ 
the one step translation  ($A\to B$),  ${\cal R}_{2\pi/3}$
(${\cal R}_{2\pi/3}^{'}$)
the three-fold rotation around a site of the
$D$ ($B$)-sublattice, and $\sigma$ the axial symmetry
keeping invariant $C$ and $D$.
$N_{el}$ is the number of elements in each class.}
\label{Table-1}
\end{table}
\begin{table}
        \begin{center}                        
                \begin{tabular}{|c|c c c c c c c c c|}
                        \hline
                        $N=32$ &&&&&&&&& \\
                        $S$ & 0 & 1 & 2 & 3 & 4 & 5 & 6 & 7 & 8 \\
                        \hline
                        $n_{\Gamma_1}(S)$ &2 & 0 & 3 & 1 & 4 & 2 & 4 & 2 & 4 \\
                        $n_{\Gamma_2}(S)$ &1 & 0 & 2 & 1 & 2 & 1 & 2 & 1 & 1 \\
                        $n_{\Gamma_3}(S)$ &3 & 0 & 5 & 2 & 6 & 3 & 6 & 3 & 5 \\
                        $n_{\Gamma_4}(S)$ &0 & 4 & 4 & 7 & 6 & 8 & 7 & 8 & 6 \\
                        $n_{\Gamma_5}(S)$ &0 & 4 & 3 & 6 & 5 & 7 & 5 & 6 & 4 \\
                        \hline
                \end{tabular}
        \end{center}
\caption[99]{Number of occurrences $n_{\Gamma_i}(S)$ of each irreducible
representation $\Gamma_i$ in $\{ ^4\tilde E\}$ as a function of  the total spin $S$. }
\label{Table-2}
\end{table}
\begin{table}
        \begin{center}                        
                \begin{tabular}{|c|c c c c c c c c c|}
                        \hline
                        $N=32$ &&&&&&&&& \\
                        $S$ & 0 & 1 & 2 & 3 & 4 & 5 & 6 & 7 & 8 \\
                        \hline
                        $n_{\Gamma_1}(S)$ &1 & 0 & 1 & 0 & 1 & 0 & 1 & 0 & 1 \\
                        $n_{\Gamma_3}(S)$ &1 & 0 & 1 & 0 & 1 & 0 & 1 & 0 & 1 \\
                        $n_{\Gamma_4}(S)$ &0 & 1 & 0 & 1 & 0 & 1 & 0 & 1 & 0 \\
                        \hline
                \end{tabular}
        \end{center}
\caption[99]{Number of occurrences $n_{\Gamma_i}(S)$ of each irreducible
representation $\Gamma_i$ in $\{ ^2\tilde E\}$ as a function of  the total spin $S$. }
\label{Table-3}
\end{table}

\end{document}